# Distinct stages of radio frequency emission at the onset of pedestal collapse in KSTAR H-mode plasmas

M H Kim[1], S G Thatipamula[2], J E Lee[1], M J Choi[2], H K Park[2,3], T Akiyama[4], and G S Yun[1*]

[1] Pohang University of Science and Technology, Pohang, Gyeongbuk 37673, Republic of Korea
[2] National Fusion Research Institute, Daejeon 34133, Republic of Korea
[3] Ulsan National Institute of Science and Technology, Ulsan 44919, Republic of Korea
[4] National Institute for Fusion Science, Toki 509-5292, Japan

E-mail: gunsu@postech.ac.kr

**Abstract**
Using a high-speed and broadband radio frequency (RF) (0.1–1 GHz) spectrum analyzer developed on the KSTAR tokamak, it is found that several distinct stages of RF emission appear at the pedestal collapse in high confinement discharges. Comparison with 2-D electron cyclotron emission (ECE) images has revealed that each stage is related to the instantaneous condition at the outboard mid-plane edge. First, high-harmonic ion cyclotron emissions (ICE) are intensified with the appearance of a non-modal filamentary perturbation in the edge within several tens of microseconds before the collapse. Then, the RF emission becomes broad toward high-frequency range (< 500 MHz) at the burst onset of the non-modal filament. During the pedestal collapse initiated by the filament burst, rapid chirping (1-3 $\mu$s) appear with additional filament bursts. The strong correlation between the RF spectra and the perturbation structure provides important clues on the stability of edge-localized modes and on the ion dynamics in the plasma boundary.

## 1. Introduction

Edge-localized modes (ELMs) are a class of spatially localized perturbations in the sharp pressure gradient region called pedestal in the edge of high confinement (*H*-mode) tokamak plasmas [1]. It is believed that the ELM triggers destruction of the edge confinement, which results in semi-periodic cycle of rapid relaxation and gradual rebuild of the pedestal. During the pedestal relaxation triggered by the ELM, the energy and particles are expelled across the plasma boundary, which can cause significant damage to plasma-facing components [2]. Therefore, it is important to understand the ELM dynamics for the safe steady-state operation in future reactors.

Many theoretical and experimental works have been carried out to investigate the dynamics of ELM evolution and pedestal collapse. Recently, the ELM dynamics has been studied using 2-D electron cyclotron emission (ECE) imaging system on the Korea Superconducting Tokamak Advanced Research (KSTAR) [3–5]. The ELM evolution typically consists of three or four different stages: initial linear growth phase of eigenmode (which is usually not clearly visible due to the background turbulent fluctuations), quiescent phase with quasi-stable eigenmode perturbation structure, transient phase with disappearance of the eigenmode, and the crash phase leading to the pedestal collapse. Just before the collapse, edge-localized non-modal filamentary perturbation emerges [5, 6] and bursts promptly leading to the pedestal collapse [5]. These observations suggest that strong nonlinearity underlies the ELM dynamics. Recently, nonlinear numerical simulations [7], theoretical models [8], experiments of ELM crash suppression with external magnetic perturbations [4, 9] suggested that the nonlinear ELM dynamics are associated with the development of **E**×**B** velocity shear across the pedestal.

Despite the significant progress in understanding the characteristics of the ELM dynamics, the crash triggering mechanism is still elusive. Owing to the complexity and fast time scale dynamics involved in the ELM crashes, novel diagnostic methods are highly desirable. For this purpose, a fast radio frequency (RF) diagnostic system has been developed on the KSTAR [10, 11].

In this paper, we report the features of dynamic changes in RF radiation (0.1–1 GHz) associated with the pedestal collapse. As will be described in section 2, the RF emission has several distinct stages around the crash time. Before the crash, the RF emission appears in several distinct harmonic lines in the spectrogram. The spacing between these lines is the deuterium cyclotron frequency at the outer mid-plane edge region. These harmonic ion cyclotron emissions (ICEs) become intensified and show frequency chirping-up/down at the onset of and during the pedestal collapse, respectively. Comparison with ECE images suggests that the dynamic changes of ICE are strongly associated with the instantaneous condition at the outboard mid-plane edge.

## 2. Experimental result

Figure 1 shows an example of *H*-mode discharge in the KSTAR shot #16176. The shot parameters are as follow: $B_0 = 1.8$ T, $I_p = 520$ kA and $W_{tot} = 330$ kJ. The total heating power is 3.8 MW with three tangential neutral beam injections (NBIs) with different pitch angles. A filter-bank spectrometer is used to measure the RF emissions with 16 different frequency bands for the whole discharge duration [10]. The last panel of Figure 1(b) represents the time traces of the 200 and 300 MHz filter-bank channels. The ECE spectrogram in the first panel of Figure 1(b) captures the lab-frame mode frequencies of the ELMs measured at the outboard mid-plane edge ($R = 217$ cm) before the pedestal collapse. In this example, the mode activities are highly dynamic showing multiple changes of mode frequency and the corresponding variations of the RF emission intensities. The ECE spectrum becomes broad at the pedestal collapse as commonly identified by the surge of the $D_\alpha$ signal. At the pedestal collapse, the RF emission maximally increases and then rapidly decreases.

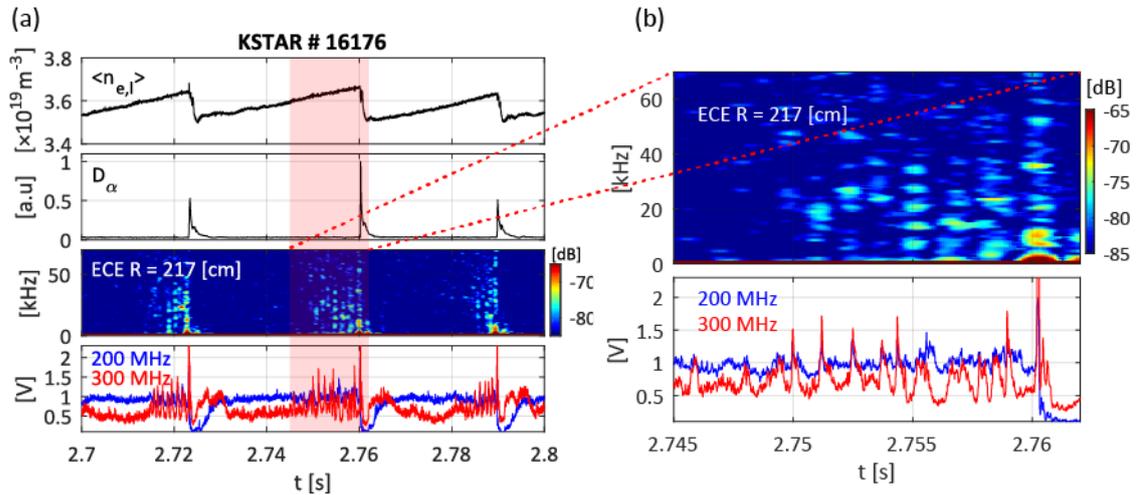

**Figure 1.** (a) Time traces of line average density, $D_\alpha$, ECE signal spectrum in outboard mid-plane edge ($R = 217$ cm), and filter-bank RF signals (200, 300 MHz) for KSTAR #16176. (b) Magnified image of ECE spectrum and filter-bank RF signals from t = 2.745 s to 2.762 s

From the filter-bank data, the trends of RF emission can be observed during multiple ELM cycles or the whole time range of discharge. KSTAR is equipped with another fast RF digitizer system (~5 GSa/s), which enables us to see the details of RF emission spectra with a high spectral resolution [11]. The high temporal (~1 $\mu$s) and spectral (~1 MHz) resolution are essential to resolve the fast ELM crash process, which proceeds within several hundreds of microseconds. Figure 2 represents the evolution of RF emission at a pedestal collapse event for KSTAR #16176. In Figure 2(a), the 200

MHz filter-bank RF channel and $D_\alpha$ are overlaid around the pedestal collapse event. The fast RF diagnostic system successfully captured the event shown in Figure 2(b). Here, $t_0$ is the time point at which the slope of the filter-bank signals changes drastically (i.e., the first time derivative is almost discontinuous). As labeled in the figure, the evolution of the RF emission can be divided into four distinct stages with respect to the time $t_0$.

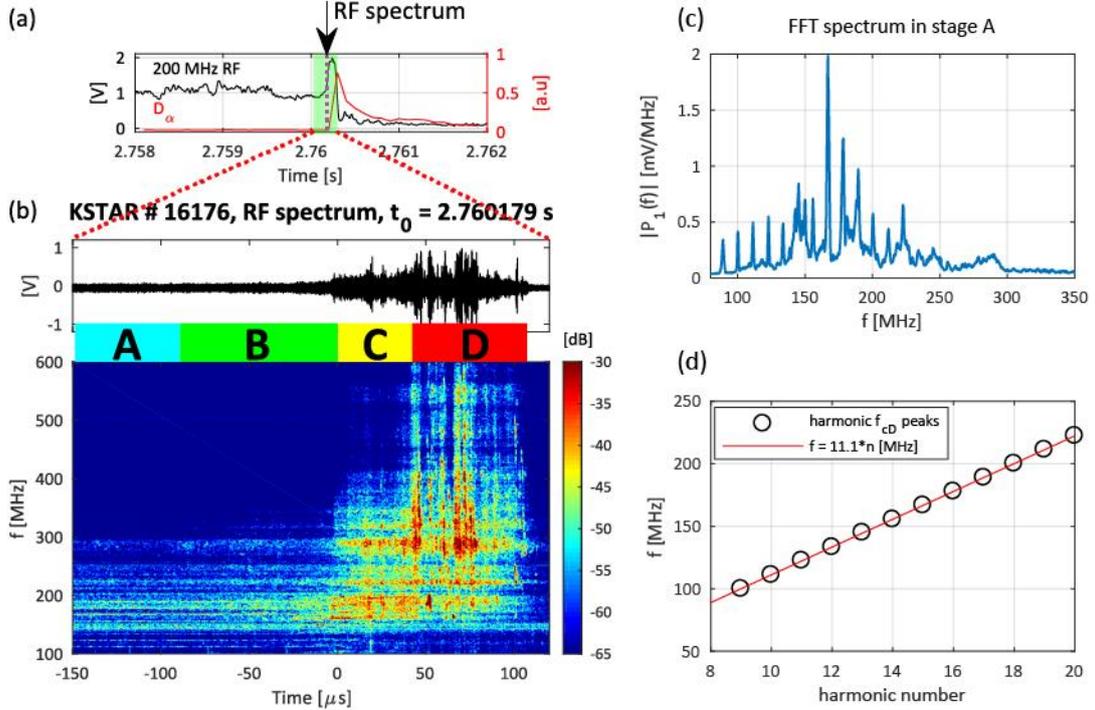

**Figure 2** (a) 200 MHz RF signal of the filter-bank spectrometer and $D_\alpha$ signal in KSTAR #16176. (b) RF bursts at the pedestal collapse. The labels (A–D) indicate the four different stages during this event. (c) Deuterium ICE harmonic peaks in stage A. (d) Linear fitting of ICE harmonic peaks in Figure 2(c) (f = 99.9–222 MHz, harmonic number, n = 9–20).

The first stage (stage A) is characterized by several line emissions in the range of 100 to 230 MHz. The spacing between these lines is 11.1±0.11 MHz (see Figure 2(c)), which corresponds to the deuterium cyclotron frequency ($f_{cD}$) at the outboard mid-plane edge ($R = 221\pm2.3$ cm). In KSTAR, it is frequently observed such discrete spectral patterns during the inter-crash period of *H*-mode discharge.

The second stage (stage B) follows from $t_0-90$ $\mu$s. The $f_{cD}$ harmonics are observed in the range of 100 to 250 MHz similar to the stage A. To emphasize the difference between stage B and A, magnified spectrogram and 10 $\mu$s averaged spectrum are compared in Figure 3(c) and (d), respectively. The main difference between these two stages stands out in the 250 to 350 MHz range. The intensities of high harmonics of $f_{cD}$ in that range are increased at the stage B. Interestingly, stage B coincides with the emergence of the non-modal structure in the outer mid-plane edge region ($R \sim 213$–$220$ cm) as illustrated by the ECE images in Figure 3(b).

In stage C from $t_0$, the RF emission is rapidly intensified as seen in the filter-bank channels in Figure 3(a). The corresponding spectrogram shows the emergence of broad emission under 500 MHz as well as the intensification of harmonic ICEs as shown in Figure 4(a). In the ECE image as illustrated in Figure 4(b), the non-modal filamentary structure starts to burst at around $t_0$ and expels the heat from the filamentary perturbation, which initiates the pedestal collapse. This suggests that the RF signals can be a good indicator for the crash onset whereas the $D_\alpha$ signal reflects only the aftermath of the collapse.

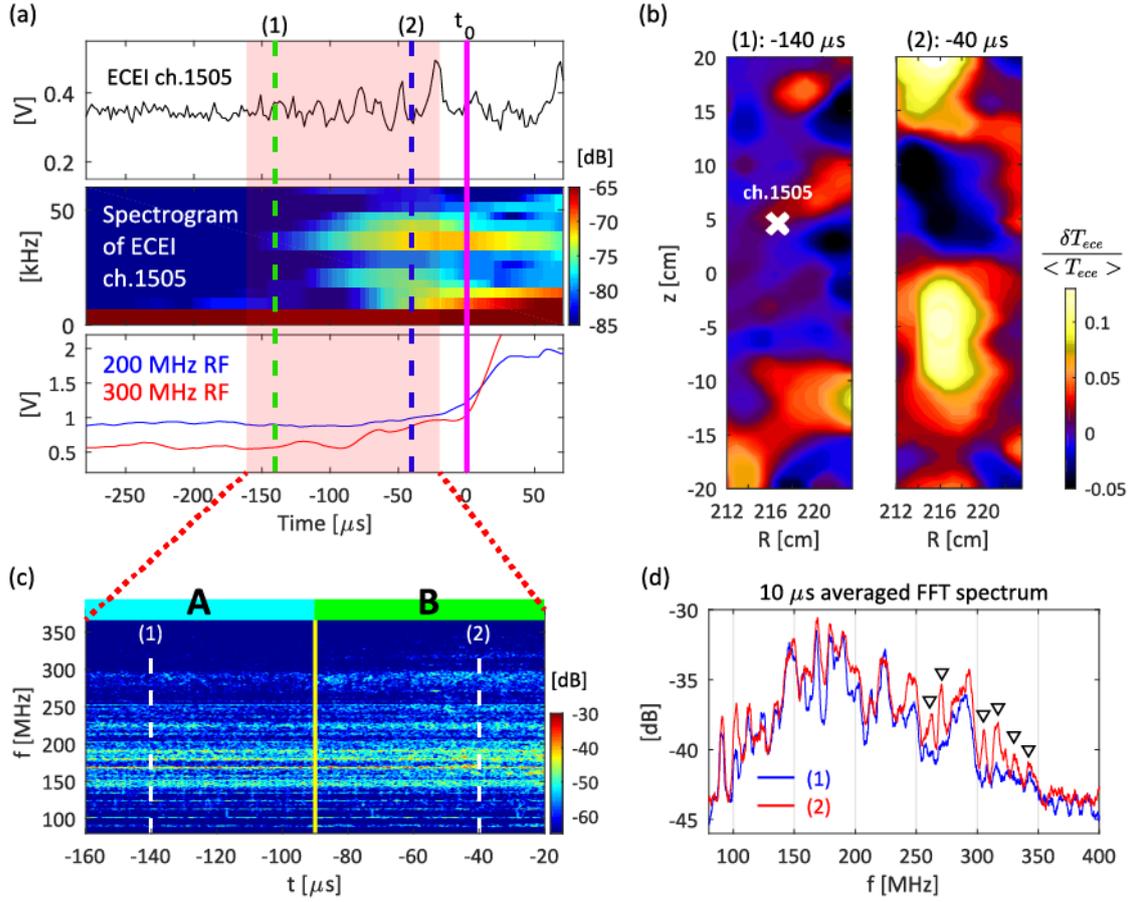

**Figure 3.** (a) ECE signal in the outer mid-plane edge ($R = 217$ cm), its spectrogram and 200, 300 MHz filter-bank RF signals at the pedestal collapse. (b) ECE images (0.5–60 kHz bandpass filtering) at (1) and (2). (c) Magnified RF spectrogram in stage A and B. (d) High-resolution spectrum graphs for 10 $\mu$s window ($\pm 5$ $\mu$s) around the times (1) and (2).

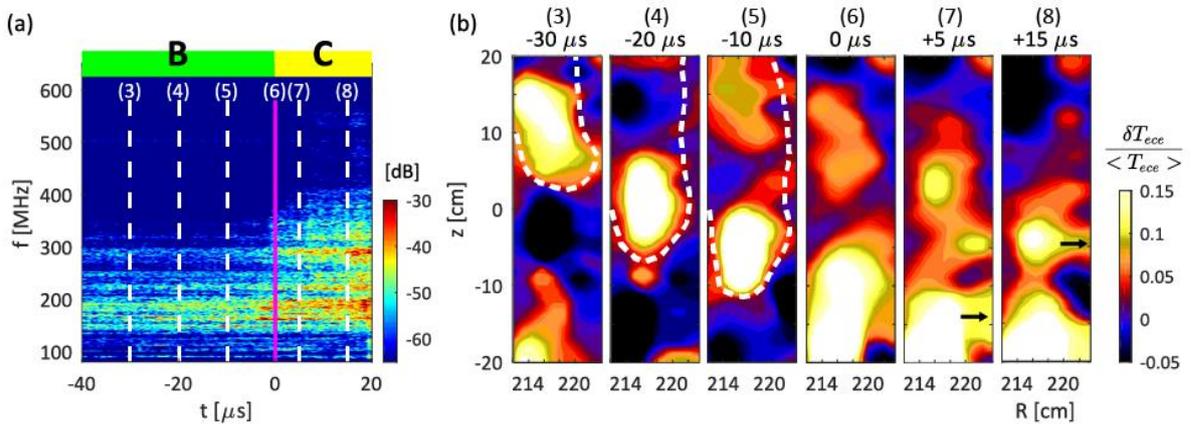

**Figure 4.** (a) Magnified RF spectrogram in stage B and C. The solid magenta line indicates $t_0$. (b) ECE images (0.5–60 kHz bandpass filtering) from the time points (3) to (8).

Following the filament burst, the pedestal collapse proceeds. During the collapse, additional filament bursts occur with 1-3 $\mu$s rapid up/down chirping in the RF spectrum as shown Figure 5. Most of these chirpings occur in step of $f_{cD}$ at the outer mid-plane edge for a broad RF range. For the case of chirpings with proton cyclotron frequency ($f_{cH}$) stepping, it was desrcibed in Ref [12].

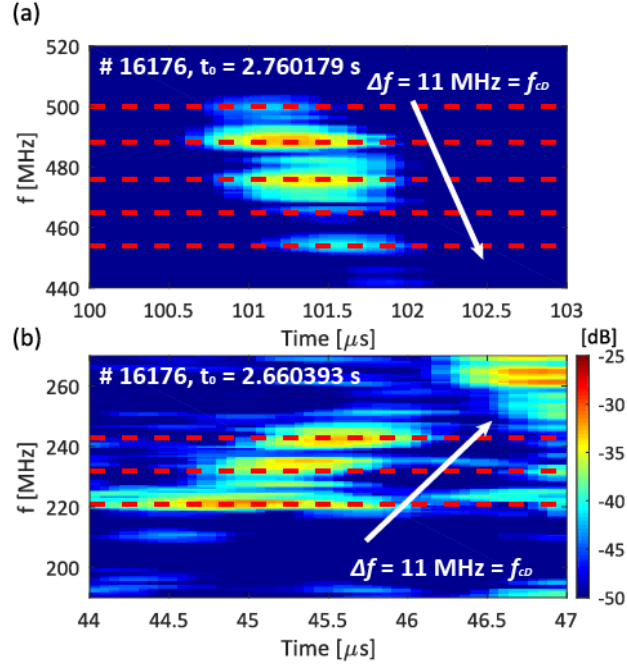

**Figure 5.** (a) RF burst with chirping down during pedestal collapse. (b) RF burst with chirping up during pedestal collapse. Here, $t_0$ is determined in the same way in Figure 2. The chirping steps in (a) and (b) are deuterium cyclotron frequency ($f_{cD}$). The KSTAR experiment used in ref. [12], they are proton cyclotron frequency ($f_{cH}$).

3. **Discussion**

In KSTAR RF observations, the uniform spacing between harmonics of ion cyclotron waves (ICWs) implies that the ICE occurs in the narrow range of the outer edge region. Previous study in JET showed similar results, which include that the ICE intensity is linearly proportional to the neutron flux, from which the population of energetic ions in the plasma edge can be inferred [13]. Based on this observation, a possible mechanism was proposed, where a subset of fusion-born energetic ions in the central plasma region can have large radial excursion to form wedge-shaped ring-type velocity distribution in the plasma edge [14]. This anisotropy in ion velocity distribution can drive the fast Alfvén-ion Bernstein waves called magneto-acoustic cyclotron instability (MCI) [14, 15]. Anisotropy in ion velocity distribution can be caused by fusion-born energetic ions or fast magnetohydrodynamic events (e.g. rapid 'tongue'-shaped deformation of flux surfaces reported in the Large Helical Device (LHD) [16]) or fast ions produced by charge exchange with NBI particles [17]. Furthermore, recent 1D3V particle-in-cell (PIC) simulation showed that the high concentrations of energetic particles can intensify the harmonic ICE [18], suggesting that the intensification in high-harmonic ICE described in Figure 3(c) and Figure 3(d) can be interpreted as the increase of fast ions at the outer edge region near the pedestal collapse. In addition, a similar 1D3V PIC simulation with time-varying local electron density reproduced the rapid stair-casing chirping of proton-ICEs observed during the pedestal collapse phase on the KSTAR [12].

On the other hand, the MCI driven by fusion-born energetic ions may not fully explain the dynamic change of RF emissions with the strong imprint of deuterium ICEs observed near pedestal collapse of the KSTAR deuterium discharges where fusion-born energetic ions are protons and tritons, implying the existence of other sources of ICEs. For the LHD, simulations of ion orbits showed that a small fraction of perpendicular neutral beam-produced fast ions can have specific orbits necessary for MCI [19]. The possibility of fast ions driven by the KSTAR tangential neutral beams [20] as a source of

MCI remains as a future research.

Our observations can suggest another possible source to enhance the high-harmonic ICWs. In KSTAR, the harmonic ICEs only appear in H-mode discharges. It is well known that inhomogeneous radial electric field ($\mathbf{E}_r$) (e.g. a well structure) is formed near the plasma periphery in H-mode discharge [21]. This suggests the inhomogeneous $\mathbf{E}_r$ structure in the pedestal can be an important factor for the generation of the deuterium ICEs. Especially, the inhomogeneity results in the strong $\mathbf{E} \times \mathbf{B}$ velocity shear in the poloidal velocity profiles [22, 23] and can be intensified near the onset of the crash [7]. As observed in Ref. [5], the nonlinear generation of axisymmetric $\mathbf{E} \times \mathbf{B}$ velocity shear can lead to the expulsion of filamentary perturbation, which can also drive the broadband RF emission like the stage C shown in Figure 4. Another relevant example may be the electromagnetic ICEs observed in the laboratory experiment in Space Physics Simulation Chamber [24] and the growth of ICWs observed in theoretical model [25, 26] showing that electromagnetic ICWs can be excited by $\mathbf{E} \times \mathbf{B}$ velocity shear. The modeling in Ref [25] and [26] considered the ionospheric plasma condition, the assumption of cold and collisionless plasma except for the motion in the parallel direction.

At the same time, the beam driven MCI [19] can be a possible candidate even for the case of tangential NBIs as in the KSTAR in the presence of strong $\mathbf{E}_r$. In the KSTAR *H*-mode discharges, substantial prompt loss (a few %) of the beam ions produced by the tangential NBIs were observed [20], implying that some fraction of the initially tangential beam ions have large orbits encompassing the low-field-side pedestal region. Those fast ions can be influenced by the strong $\mathbf{E}_r$, progressively acquiring perpendicular momentum as they circle around the torus, and subsequently drive the MCI.

4. **Summary**

The relationship between the RF spectra and the state of the plasma edge region has been identified using the fast RF and imaging diagnostics. The RF emission near the ELM crash evolves in distinct stages: intensified high harmonics in deuterium ICE, broad emission toward high-frequency range at the onset of the crash, and broadband burst and rapid frequency chirping-up/down during the pedestal collapse. Interpretation of RF evolution raises many questions: the mechanism for generating high harmonics of deuterium ICE and for driving the rapid chirping-up/down event. The interpretation based on MCI can partially explain some of these phenomena. The coincidence of ICE intensification and non-modal perturbation structure in the outer mid-plane edge suggests another important factor in ICE generation, the velocity shear driven by radial electric field well. The RF observation suggests that ICE can play an important role in diagnosing the ion dynamics during the ELM crash.

**Acknowledgements**
We thank Mr. Hee–Dong Kim in the Keysight Technologies Korea and the KSTAR team for technical support. This work was supported by the NRF Korea under grant No. NRF-2017M1A7A1A03064231 and BK21+ program.